\documentclass{optica-article}

\journal{opticajournal} 

\articletype{Research Article}

\begin{document}

\title{A surface-normal photodetector as nonlinear activation function in diffractive optical neural networks}

\author{Farshid Ashtiani,\authormark{1,2,*} Mohamad Hossein Idjadi\authormark{1,2}, Ting-Chen Hu,\authormark{1} Stefano Grillanda,\authormark{1} David Neilson,\authormark{1} Mark Earnshaw,\authormark{1} Mark Cappuzzo,\authormark{1} Rose Kopf,\authormark{1} Alaric Tate,\authormark{1} and Andrea Blanco-Redondo\authormark{1}}

\address{\authormark{1}Nokia Bell Labs, 600 Mountain Ave, Murray Hill, NJ 07974, USA\\
\authormark{2}These authors contributed equally to this work.}

\email{\authormark{*}farshid.ashtiani@nokia-bell-labs.com} 


\begin{abstract*} 
Optical neural networks (ONNs) enable high speed parallel and energy efficient processing compared to conventional digital electronic counterparts. However, realizing large scale systems is an open problem. Among various integrated and non-integrated ONNs, free-space diffractive ONNs benefit from a large number of pixels of spatial light modulators to realize millions of neurons. However, a significant fraction of computation time and energy is consumed by the nonlinear activation function that is typically implemented using a camera sensor. Here, we propose a novel surface-normal photodetector (SNPD) with a nonlinear response to replace the camera sensor that enables about three orders of magnitude faster (5.7 $\mu$s response time) and more energy efficient (less than 10 nW/pixel) response. Direct efficient vertical optical coupling, polarization insensitivity, inherent nonlinearity with no control electronics, low optical power requirements, and the possibility of implementing large scale arrays make the SNPD a promising nonlinear activation function for diffractive ONNs. To show the applicability, successful classification simulation of MNIST and Fashion MNIST datasets using the measured response of SNPD with accuracy comparable to that of an ideal ReLU function are demonstrated. 

\end{abstract*}

\section{Introduction}

As artificial neural networks are more widely utilized in a variety of applications from pattern recognition \cite{pattern1,pattern2} to medical diagnosis \cite{med1,med2}, there is an increasing need for faster and more energy efficient hardware platforms. Optical neural networks (ONNs) benefit from massive parallelism and different multiplexing schemes, such as wavelength, mode, time, and polarization, to enable processing with high energy efficiency at the speed of light \cite{PhotonANN}. Hence, various ONN implementations have been demonstrated both using bench-top setups \cite{mainref,bench2,bench3} as well as integrated platforms that enable smaller size and higher energy efficiency \cite{integ1,integ2,integ3}. 

\begin{figure}[t]
\centering{\includegraphics[width=\linewidth]{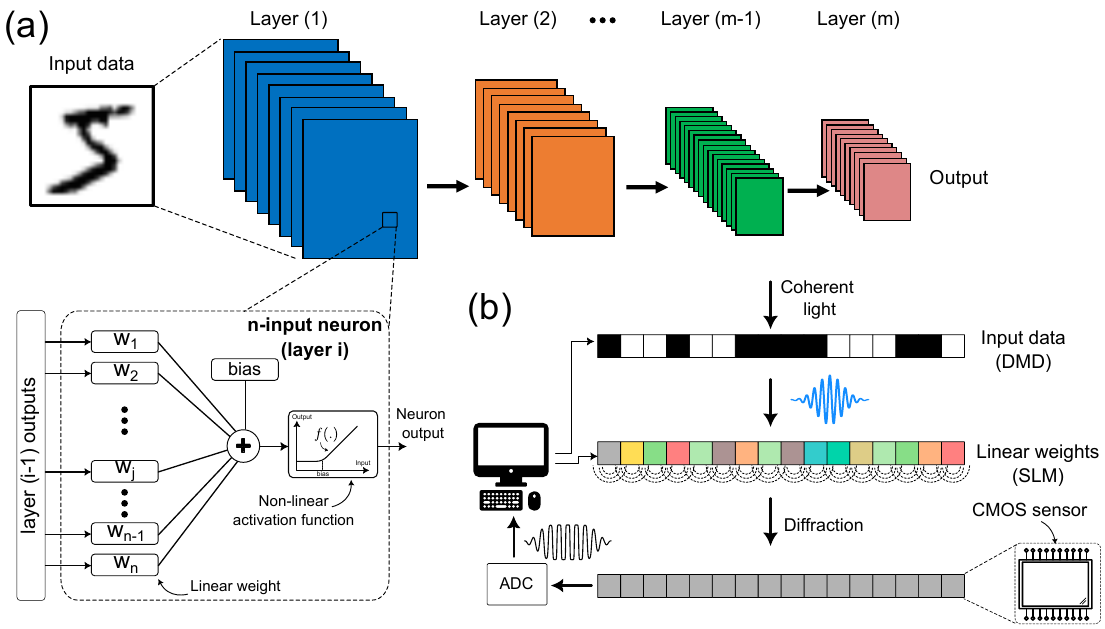}}
\caption{(a) Typical feed-forward neural network architecture with multiple layers of interconnected neurons. The neural output is generated by passing the weighted-sum of the inputs through a nonlinear activation function. (b) Diffractive ONN architecture using a DMD to generate the input signals and SLM to apply corresponding weights to the inputs \cite{mainref}. Conventionally, a CMOS sensor acts as a detector and/or nonlinear activation function.}
\label{ONN}
\end{figure}

Despite the significant progress, scaling ONNs to thousands or millions of neurons and multiple layers to perform more complex tasks, is one of the main issues that integrated ONNs face \cite{mainref}. Complex and area-consuming photonic routing in commercially available platforms, larger on-chip propagation loss, and intricate electronic control circuitry to compensate for fabrication-induced errors, result in lower energy efficiency, packaging complexities, and impractically large integrated systems.

Free-space diffractive ONNs, on the other hand, enable orders of magnitude larger number of neurons compared to integrated ONNs, as well as more flexibility to implement different network configurations \cite{mainref,bench2}. Such systems are especially useful for image and video processing and classification as they directly process the input pictures or video frames with large number of pixels. Figure \ref{ONN}(a) shows the conceptual schematic of a feed-forward neural network with multiple layers of neurons, where each neuron performs linear (weight and sum) and nonlinear (activation function) computations on its inputs. Correspondingly, a diffractive ONN architecture that performs the linear and nonlinear computations is shown in Fig. \ref{ONN}(b). A laser source illuminates a digitally controlled micro-mirror device (DMD) that modulates the intensity of the incoming light with the input data to the network. A spatial light modulator (SLM) is used to implement linear weights. Large number of pixels of commercially available SLMs enable ONNs with millions of neurons per layer. The diffracted signals from the SLM are then directed towards the camera to apply the nonlinear activation function on the weighted-sum of the inputs. So far, the nonlinear activation function has been implemented either digitally after forming the image on a camera \cite{bench2}, or using the inherent nonlinear photoelectric response of the CMOS sensor \cite{mainref}. In either case, the total computation time is mainly limited by the sensor exposure time which for commercial cameras is several milliseconds. For instance, in Ref\cite{mainref}, despite achieving an impressive performance of more than 200 tera operations per second (TOPS) and more than 1 TOPS/W, about 64\% of the total processing time and 15\% of total power consumption (about 6 $\mu$W per pixel) are consumed by the sensor. Therefore, a faster and more energy efficient implementation of the nonlinear activation function can significantly improve the computation speed and energy efficiency of such systems. Note that in most diffractive ONNs only one neural layer is implemented using this setup and the full neural network is realized by re-using the same architecture but with different parameters. The output of the layer is always in electrical domain that drives the DMD after some processing. Therefore, an optical-in electrical-out (O-E) nonlinearity would best fit such systems.  

Here we propose a novel implementation of the nonlinear activation function using a surface-normal nonlinear photodetector (SNPD) to significantly improve the speed and energy efficiency of a diffractive ONN. The SNPD is formed by a vertical p-i-n structure contained in a Fabry-Perot cavity. These devices have been used previously as high-speed electro-optic modulators operating according to the quantum confined Stark effect \cite{SNEAM1,SNEAMPHYS,MillerSNEAMs,SNEAMlowvoltage,scaling}. However, light coupled to these devices generates a photocurrent \cite{MillerSNEAMs}, and hence they can be used as photodetectors as well. Also, under high light intensity, nonlinearities induced by thermal effects arise. In this work, we use the nonlinear behavior of the SNPD photocurrent as a function of the incident optical power to realize a nonlinear activation function as an improved alternative to the camera sensor. The SNPD is a polarization-independent device and light can be vertically coupled to it with a high efficiency and without any additional coupling devices that ease its deployment within a free space ONN setup. In this work, we show that a reverse-biased SNPD (\textit{i.e.,} each pixel) has a response time of about 5.7 $\mu$s (3-dB bandwidth of 61 kHz) while consuming less than 10 nW of static power that make it about three orders of magnitude faster and more energy efficient than commercially available camera sensors. As a result, the activation function will not be a performance bottleneck of the system. As a proof of concept, the measured characteristics of SNPD is used in a neural network simulation platform to classify MNIST and Fashion MNIST datasets. In these tests, accuracies of 97\% and 89\% are achieved, respectively, showing a performance comparable with that of a standard rectified linear unit (ReLU) activation function. Note that the SNPD is primarily proposed to be utilized in a diffractive ONN setup as an O-E nonlinearity. Other solutions such as all-optical \cite{allopt1,allopt2,allopt3} and O-E-O \cite{OEO1} require additional coupling devices (\textit{e.g.}, grating couplers), polarization control, additional photodetectors to generate an electrical output, larger size per pixel, and control electronic circuitry to realize nonlinearity (especially in the case of O-E-O) that result in more complexity and less energy efficiency and make scaling more challenging. Although they enable faster response time than the SNPD, due the millisecond-scale response time of the SLM, the performance of the overall system will not improve and this only results in more energy consumption. Therefore, SNPD best fits a diffractive ONN setup.


\section{SNPD structure and characterization}


\begin{figure}[t]
\centering{\includegraphics[width=11cm]{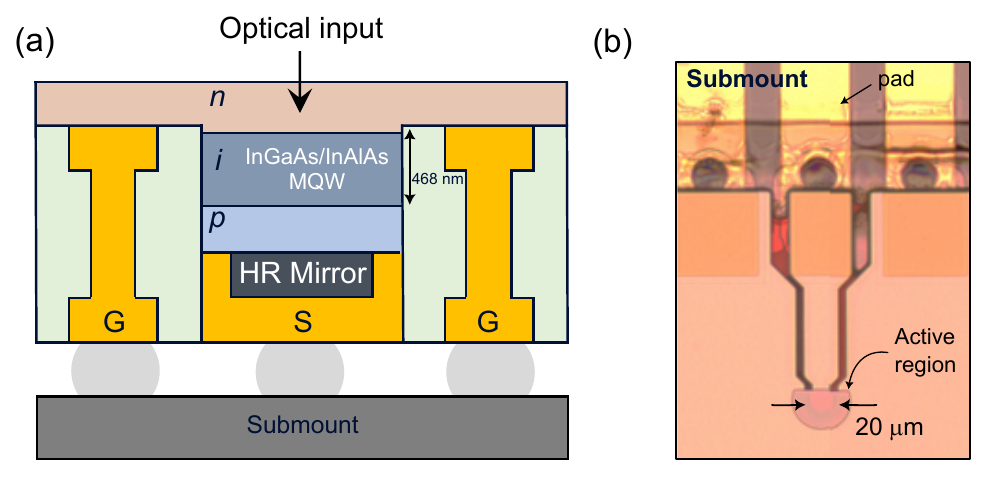}}
\caption{(a) Sketch of the cross-section of a SNPD. (b) Top-view photograph of a SNPD with 20 $\mu$m active area diameter.}
\label{SNEAM_structure}
\end{figure}

Figure \ref{SNEAM_structure}(a) shows a sketch of the cross-section of the SNPD used in this work. It is composed of a multi-quantum-well (MQW) stack placed in the intrinsic region of a vertical \textit{p-i-n} structure. The MQW is formed by 36 periods of In$_{0.53}$Ga$_{0.47}$As wells with 9 nm thickness and In$_{0.52}$Al$_{0.48}$As barriers with 4 nm thickness. The total thickness of the MQW is 468 nm, which is equivalent to one wavelength at about 1540 nm. The \textit{p-i-n} stack is then inserted in an asymmetric Fabry-Perot resonant cavity with a high-reflectivity (HR) mirror on the bottom of the structure and a partial reflectivity top mirror formed by the semiconductor/air interface. Other MQWs with different composition (such as Si/SiGe) \cite{MillerSNEAMs} and thickness \cite{SNEAMlowvoltage} may be used as well in such a structure. The SNPD used in this work has active area diameter of 20 $\mu$m. The top-view microphotograph of the device is shown in Fig. \ref{SNEAM_structure}(b). The chip is bonded to a submount with single-ended ground-signal-ground metal pads that allow application of an electric field orthogonal to the layers of the MQW region. Note that devices with smaller active area can be designed in order to reduce the form factor when placed within an array \cite{SNEAM1}. The details of the fabrication process are described in Ref\cite{SNEAMlowvoltage}.

Typically, when used as a modulator, such a device operates according to the quantum confined Stark effect: upon application of a reverse bias voltage, the MQW absorption edge shifts in wavelength and produces amplitude modulation of the optical output signal. In this work, while we still apply a reverse bias voltage, we use it as a photodetector and work at wavelengths much longer than those typically used for modulation. 

\begin{figure}[t]
\centering{\includegraphics[width=11cm]{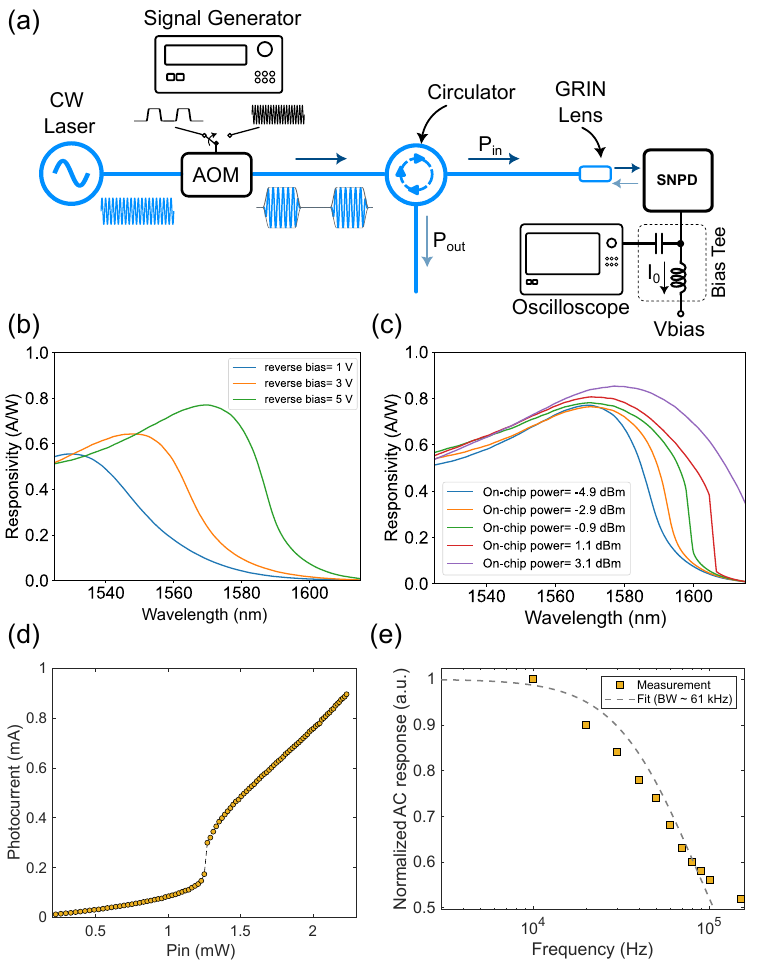}}
\caption{(a) Experimental setup used to characterize the device. (b) SNPD responsivity as a function of optical wavelength for different reverse bias voltages and an on-chip optical power of -4.9 dBm. In this case, the AOM is bypassed. (c) SNPD responsivity for different input optical powers showing a nonlinear behavior at wavelengths longer than 1580 nm. Here too, the AOM is bypassed. (d) SNPD photocurrent as a function of the input optical power (P$_{in}$) measured at the wavelength 1598 nm. The modulation signal is turned off in this measurement. (e) frequency response of the SNPD measured at the wavelength of 1598 nm, while the AOM modulates the input optical signal.}
\label{Measurement}
\end{figure}

Figure \ref{Measurement}(a) shows the experimental setup to characterize the SNPD in the linear and nonlinear regions. The output light of a tunable continuous wave (CW) laser is coupled orthogonally to the surface of the SNPD chip using a standard single mode fiber and a GRIN lens. The GRIN lens is used to reimage the optical mode of the standard fiber on the SNPD top surface with about 80\% coupling efficiency while allowing to move the fiber farther away from the chip, but does not change the mode size. A fiber-optic based circulator allows to separate light at the input and output of the SNPD. Note that the reflected optical signal (P$_{out}$ in Fig. \ref{Measurement}(a)) is used when the device operates in the modulator mode. As mentioned before, there is no need for any optical polarization control of the light as the device is fully polarization independent \cite{SNEAM1}. Moreover, the SNPD is placed on a thermo-electric cooler (TEC) to stabilize the working temperature of the device. Due to a broad wavelength range of operation of the SNPD \cite{SNEAM1}, no complex and power hungry closed-loop wavelength locking mechanism is necessary. To later characterize the nonlinear response time of the SNPD, an acousto-optic modulator (AOM) is driven with a 27 MHz CW signal by an arbitrary signal generator. In this mode of operation, the AOM only frequency shifts the laser with an insertion loss of about 3.5 dB.

In the first experiment, the responsivity of the SNPD in the linear region (\textit{i.e.}, low optical power) as a function wavelength and for different reverse bias conditions is measured. In this case, the AOM is bypassed and no amplitude modulation is performed. Figure \ref{Measurement}(b) shows the responsivity of the SNPD as a function of optical wavelength for three different reverse bias voltages and a fixed on-chip optical power of -4.9 dBm (estimated after de-embedding the loss of other components). As the reverse bias value increases, the absorption edge red-shifts, resulting in higher a peak peak responsivity. To achieve a high photocurrent, a reverse bias voltage of 5 V is used in all of the following experiments. 

In the second experiment and to study the nonlinear behavior of the SNPD as the input optical power changes, the responsivity of the device for a reverse bias voltage of 5 V and different input optical power values is measured. As shown in Fig. \ref{Measurement}(c), for optical wavelengths shorter than 1580 nm, the responsivity graphs for different input optical powers are similar and no significant nonlinearity is observed. However, for longer wavelengths, as the input optical power increases, the difference between the responsivity graphs becomes more significant, showing the nonlinear behavior of the SNPD. This behavior is dominated by thermal effects \cite{SNEAMPHYS} and once the optical power exceeds a certain threshold for a given wavelength, the generated photocurrent increases at a higher rate, resulting in a larger responsivity. 

The third experiment is performed to characterize the nonlinear response of the SNPD that is to be used in an ONN. The laser wavelength is fixed to 1598 nm while the optical power is swept. As shown in Fig. \ref{Measurement}(d), the photocurrent is a nonlinear function of the input optical power which resembles a ReLU function at 1598 nm. For optical power of larger than 1.25 mW, the change in the photocurrent significantly increases. The measured characteristic is later used in a neural network to confirm its applicability as a nonlinear activation function. Note that the threshold power is a function of the cavity design and biasing conditions and can be adjusted. Moreover, the optical wavelength of 1598 nm is chosen as it results in a close approximation of the ReLU nonlinear response. However, other wavelengths can be selected depending on the application and the desired type of nonlinear function. Since one neural layer is typically implemented using diffractive ONNs, the laser power can be set properly to maintain a sufficient optical power level at the SNPD to trigger the nonlinearity.

To measure the bandwidth of the SNPD in the proposed mode of operation, a train of square wave pulses is applied to the AOM to amplitude modulate the CW laser with an extinction ratio of greater than 35 dB. Note that the amplitude of the modulation signal is large enough to switch the input optical power between less than 1 $\mu$W and a value larger than the threshold power which is about 1.25 mw. This way, we emulate a large change in the weighted-sum signal to find the worst case scenario for the response time. In this experiment, the modulation frequency is varied and the amplitude of the AC voltage is measured across a 50 $\Omega$ load on an oscilloscope. Figure \ref{Measurement}(e) shows the normalized AC response of the SNPD (yellow squares) where fitting a single pole transfer function suggests a 3-dB bandwidth of 61 kHz, that is equivalent to a rise time (response time) of about 5.7 $\mu$s. This is about three orders of magnitude faster than the typical millisecond response time of camera sensors.


\section{Neural network simulation results}

\begin{figure}[t]
\centering{\includegraphics[width=11cm]{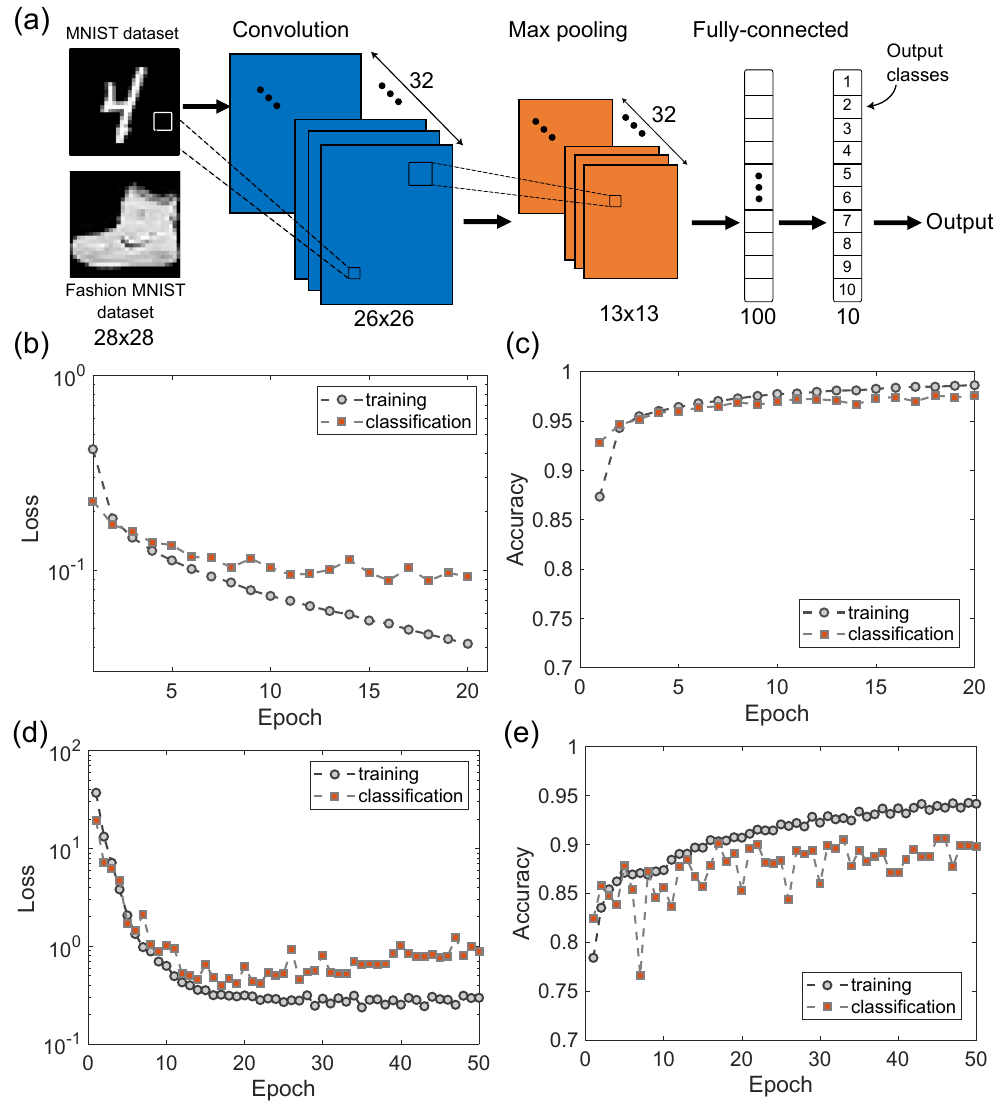}}
\caption{MNIST and Fashion MNIST data classification. (a) Architecture of the neural network used in this work. (b) Cross entropy loss and (c) classification accuracy as a function of number of epochs, showing the results both for training and test for the MNIST dataset. (d) Cross entropy loss and (e) classification accuracy as a function of number of epochs, showing the results both for training and test for Fashion MNIST dataset.}
\label{class}
\end{figure}
To demonstrate the applicability of the nonlinear response of the SNPD in a neural network, the measured transfer function of the device in Fig. \ref{Measurement}(d) is used as the activation function in a neural network simulation platform to classify MNIST and Fashion MNIST datasets. Figure \ref{class}(a) shows the architecture of a simple neural network used in this work. The $28 \times 28$-pixel images are input to the convolution layer with 32 parallel $3 \times 3$ kernels with a stride (step size) of one and SNPD response as its activation function that replaces the standard ReLU function. A maxpooling layer down-samples the output of the convolution layer and is followed by a fully-connected layer with 100 neurons and SNPD response as the nonlinearity. Finally, 10 neurons with softmax activation generate the classification results of the network. The neural network is implemented using Tensorflow libraries. Stochastic gradient descent with a learning rate of 0.01 and momentum of 0.9 is used as the optimizer and with a categorical cross-entropy as the loss function. The input images are fed to the network with a batch size of 32. Moreover, random normal kernel initialization is used throughout the network. 

Figures \ref{class}(b) and \ref{class}(c) show the training and test cross-entropy loss and classification accuracy, respectively, both as a function of the number of epochs. Using the measured SNPD nonlinear response, the network achieves a test classification accuracy of about 97$\%$. As a reference, the same network with the standard ReLU function achieves the same accuracy. 

In the second test, the same network is used to classify the Fashion MNIST dataset consisting of $28 \times 28$-pixel images of 10 different types of clothing (Fig. \ref{class}(a)). While the network architecture is the same, Adam optimizer is used instead of stochastic gradient descent for faster convergence. Moreover, He Uniform is used for kernel initialization. Figures \ref{class}(d) and \ref{class}(e) show the training and test loss and accuracy as a function of the number of epochs, respectively. An accuracy of about 88.5$\%$ is achieved while the same network with ReLU function achieves about 89$\%$. Note that the lower accuracy compared to the MNIST classification case is due to more complex features in the Fashion MNIST dataset and can be improved by using a network that is better optimized for this application.

\section{Discussion and summary}
It should be noted that the proposed SNPD as the nonlinear activation function can be scaled to one-dimensional (1D) and two-dimensional (2D) arrays with large number of devices (pixels), similar to camera sensors. For instance, in Ref\cite{scaling}, a $288 \times 132$ array of similar devices is demonstrated. Therefore, high-resolution 2D array of nonlinear activation functions can be used in a diffractive ONN. 
 
In summary, we demonstrated the applicability of a surface-normal nonlinear photodetector in free-space diffractive ONNs to realize the O-E neural nonlinearity as an alternative to the commonly used camera sensors. Significantly faster response time of 5.7 $\mu$s removes the nonlinear activation function as a computation time bottleneck. The reverse biased SNPD consumes less than 10 nW of static power per pixel which in turn improves the overall energy efficiency of an ONN. Moreover, polarization-independent operation of the SNPD together with direct optical coupling and the possibility of implementing large-scale 1D and 2D arrays of the device, make it a promising candidate to be used in a free space ONN setup.

\section{Backmatter}

\begin{backmatter}

\bmsection{Disclosures} The authors declare no conflicts of interest.

\end{backmatter}

\bibliography{sample}

\begin{thebibliography}{10}
\newcommand{\enquote}[1]{``#1''}

\bibitem{pattern1}
T.~Serre, L.~Wolf, S.~Bileschi, M.~Riesenhuber, and T.~Poggio, \enquote{Robust
  object recognition with cortex-like mechanisms,} {\protect\JournalTitle{IEEE
  Transactions on Pattern Analysis and Machine Intelligence}} \textbf{29},
  411--426 (2007).

\bibitem{pattern2}
D.~Wang, J.~Su, and H.~Yu, \enquote{Feature extraction and analysis of natural
  language processing for deep learning english language,}
  {\protect\JournalTitle{IEEE Access}} \textbf{8}, 46335--46345 (2020).

\bibitem{med1}
M.~Daniali, D.~D. Salvucci, and M.~T. Schultheis, \enquote{Understanding driver
  behavior after concussion: A machine-learning approach,}
  {\protect\JournalTitle{Proceedings of the Human Factors and Ergonomics
  Society Annual Meeting}} \textbf{64}, 1911--1915 (2020).

\bibitem{med2}
B.~Yuan, D.~Yang, B.~E.~G. Rothberg, H.~Chang, and T.~Xu, \enquote{Unsupervised
  and supervised learning with neural network for human transcriptome analysis
  and cancer diagnosis,} {\protect\JournalTitle{Scientific Reports}}
  \textbf{10}, 19106 (2020).

\bibitem{PhotonANN}
Q.~Zhang, H.~Yu, M.~Barbiero, B.~Wang, and M.~Gu, \enquote{Unsupervised and
  supervised learning with neural network for human transcriptome analysis and
  cancer diagnosis,} {\protect\JournalTitle{Light: Science and Applications}}
  \textbf{8}, 42 (2019).

\bibitem{mainref}
T.~Zhou, X.~Lin, J.~Wu, Y.~Chen, H.~Xie, Y.~Li, J.~Fan, H.~Wu, L.~Fang, and
  Q.~Dai, \enquote{Large-scale neuromorphic optoelectronic computing with a
  reconfigurable diffractive processing unit,} {\protect\JournalTitle{Nature
  Photonics}} \textbf{15}, 367–373 (2021).

\bibitem{bench2}
J.~Chang, V.~Sitzmann, X.~Dun, W.~Heidrich, and G.~Wetzstein, \enquote{Hybrid
  optical-electronic convolutional neural networks with optimized diffractive
  optics for image classification,} {\protect\JournalTitle{Scientific Reports}}
  \textbf{8}, 12324 (2018).

\bibitem{bench3}
Y.~Zuo, B.~Li, Y.~Zhao, Y.~Jiang, Y.-C. Chen, P.~Chen, G.-B. Jo, J.~Liu, and
  S.~Du, \enquote{All-optical neural network with nonlinear activation
  functions,} {\protect\JournalTitle{Optica}} \textbf{6}, 1132--1137 (2019).

\bibitem{integ1}
F.~Ashtiani, A.~J. Geers, and F.~Aflatouni, \enquote{An on-chip photonic deep
  neural network for image classification,} {\protect\JournalTitle{Nature}}
  \textbf{606}, 501–506 (2022).

\bibitem{integ2}
J.~Feldmann, N.~Youngblood, M.~Karpov, H.~Gehring, X.~Li, M.~Stappers, M.~L.
  Gallo, X.~Fu, A.~Lukashchuk, A.~S. Raja, J.~Liu, C.~D. Wright, A.~Sebastian,
  T.~J. Kippenberg, W.~H.~P. Pernice, and H.~Bhaskaran, \enquote{Parallel
  convolutional processing using an integrated photonic tensor core,}
  {\protect\JournalTitle{Nature}} \textbf{589}, 52–58 (2021).

\bibitem{integ3}
A.~N. Tait, T.~F. de~Lima, E.~Zhou, A.~X. Wu, M.~A. Nahmias, B.~J. Shastri, and
  P.~R. Prucnal, \enquote{Neuromorphic photonic networks using silicon photonic
  weight banks,} {\protect\JournalTitle{Scientific Reports}} \textbf{7}, 7430
  (2017).

\bibitem{SNEAM1}
S.~Grillanda, T.-C. Hu, D.~Neilson, N.~Basavanhally, Y.~Low, H.~Safar,
  M.~Cappuzzo, R.~Kopf, A.~Tate, G.~Raybon, A.~Adamiecki, N.~Fontaine, and
  M.~Earnshaw, \enquote{107 gb/s ultra-high speed, surface-normal
  electroabsorption modulator devices,} {\protect\JournalTitle{Journal of
  Lightwave Technology}} \textbf{38}, 804--810 (2020).

\bibitem{SNEAMPHYS}
S.~Grillanda, T.-C. Hu, D.~Neilson, and M.~Earnshaw, \enquote{Power insensitive
  surface-normal electroabsorption modulators,} {\protect\JournalTitle{Opt.
  Lett.}} \textbf{45}, 4472--4475 (2020).

\bibitem{MillerSNEAMs}
R.~M. Audet, E.~H. Edwards, K.~C. Balram, S.~A. Claussen, R.~K. Schaevitz,
  E.~Tasyurek, Y.~Rong, E.~I. Fei, T.~I. Kamins, J.~S. Harris, and D.~A.~B.
  Miller, \enquote{Surface-normal ge/sige asymmetric fabry–perot optical
  modulators fabricated on silicon substrates,} {\protect\JournalTitle{Journal
  of Lightwave Technology}} \textbf{31}, 3995--4003 (2013).

\bibitem{SNEAMlowvoltage}
S.~Grillanda, T.-C. Hu, D.~Neilson, and M.~Earnshaw, \enquote{Low-voltage
  surface-normal electroabsorption modulators,} {\protect\JournalTitle{Opt.
  Lett.}} \textbf{46}, 5425--5428 (2021).

\bibitem{scaling}
U.~Arad, E.~Redmard, M.~Shamay, A.~Averboukh, S.~Levit, and U.~Efron,
  \enquote{Development of a large high-performance 2-d array of gaas-algaas
  multiple quantum-well modulators,} {\protect\JournalTitle{IEEE Photonics
  Technology Letters}} \textbf{15}, 1531--1533 (2003).

\bibitem{allopt1}
Y.~Shi, J.~Ren, G.~Chen, W.~Liu, C.~Jin, X.~Guo, Y.~Yu, and X.~Zhang,
  \enquote{Nonlinear germanium-silicon photodiode for activation and monitoring
  in photonic neuromorphic networks,} {\protect\JournalTitle{Nature
  Communications}} \textbf{13} (2022).

\bibitem{allopt2}
B.~Wu, H.~Li, W.~Tong, J.~Dong, and X.~Zhang, \enquote{Low-threshold
  all-optical nonlinear activation function based on a ge/si hybrid structure
  in a microring resonator,} {\protect\JournalTitle{Opt. Mater. Express}}
  \textbf{12}, 970--980 (2022).

\bibitem{allopt3}
Y.~Zuo, B.~Li, Y.~Zhao, Y.~Jiang, Y.-C. Chen, P.~Chen, G.-B. Jo, J.~Liu, and
  S.~Du, \enquote{All-optical neural network with nonlinear activation
  functions,} {\protect\JournalTitle{Optica}} \textbf{6}, 1132--1137 (2019).

\bibitem{OEO1}
M.~M.~P. Fard, I.~A.~D. Williamson, M.~Edwards, K.~Liu, S.~Pai, B.~Bartlett,
  M.~Minkov, T.~W. Hughes, S.~Fan, and T.-A. Nguyen, \enquote{Experimental
  realization of arbitrary activation functions for optical neural networks,}
  {\protect\JournalTitle{Opt. Express}} \textbf{28}, 12138--12148 (2020).

\end{thebibliography}


\end{document}